\documentclass[trackchanges,twocolumn,twocolappendix]{aastex701}

\usepackage{url}
\usepackage[whole]{bxcjkjatype}
\usepackage{comment}
\usepackage{bm}
\usepackage{physics}
\usepackage{color}
\usepackage{soul}
\usepackage{siunitx}

\def\revise#1#2{#2}

\begin{document}

\title{Multi-hierarchy simulation of Riemann problem for reconnection exhausts}

\author[orcid=0009-0005-4643-1660,gname='Keita',sname='Akutagawa']{Keita Akutagawa}
\affiliation{Department of Earth and Planetary Science, Graduate School of Science, The University of Tokyo, Tokyo, Japan}
\email[show]{akutagawa@eps.s.u-tokyo.ac.jp}  

\author[orcid=0000-0001-7891-3916,gname=Shinsuke,sname='Imada']{Shinsuke Imada} 
\affiliation{Department of Earth and Planetary Science, Graduate School of Science, The University of Tokyo, Tokyo, Japan}
\email{imada@eps.s.u-tokyo.ac.jp}  

\author[orcid=0000-0002-7136-8190,gname=Munehito,sname='Shoda']{Munehito Shoda} 
\affiliation{Department of Earth and Planetary Science, Graduate School of Science, The University of Tokyo, Tokyo, Japan}
\email{shoda@eps.s.u-tokyo.ac.jp}  

\begin{abstract}
Magnetic reconnection drives a wide range of astrophysical plasma phenomena, including solar flares, by converting magnetic energy into plasma energy through changes in magnetic field topology. Petschek reconnection is a magnetohydrodynamic (MHD) model in which magnetic field lines reconnect within a localized diffusion region, and a pair of switch-off slow shocks forms outside this region, enabling efficient energy conversion. Whether this picture remains valid when kinetic effects are included remains an open question. In this study, we examine the formation and properties of slow shocks associated with reconnection exhausts by solving a two-dimensional Riemann problem using a multi-hierarchy framework that couples MHD and particle-in-cell (PIC) simulations. We find that a slow shock close to the switch-off limit forms in the MHD domain even when slow shock formation is suppressed in the PIC domain, and that this behavior is insensitive to the size of the PIC domain. The formation of the slow shock further promotes plasma isotropization within the PIC domain. These results suggest that Petschek-like reconnection remains viable \revise{in weakly collisional systems where the mean free path is small compared to the global system size even when kinetic effects are present locally, such as solar flares.}{in collisionless-collisional systems, such as solar flares, where temperature anisotropy appears to be relaxed far from the reconnection region.}
\end{abstract}

\keywords{\uat{Astronomical simulations}{1857} --- \uat{Solar magnetic reconnection}{1504} --- \uat{Planetary magnetospheres}{997} \uat{Solar flares}{1496} --- \uat{Space plasmas}{1544}}


\section{Introduction}\label{chap1:introduction}

Magnetic reconnection is a fundamental process that converts magnetic energy into plasma energy and plays important roles across a wide range of astrophysical environments, including solar flares and substorms in the Earth's magnetosphere \citep{giovanelli1946, dungey1961}. Many observations show that the timescale of the solar flares is several hours, which is much faster than predicted value by the classical Sweet-Parker model \citep{sweet1958, parker1957}. Although various models of fast magnetic reconnection have been proposed, no single unified model has yet been established. This is because, while the signatures of magnetic reconnection are observed on magnetohydrodynamic (MHD) scale, kinetic effects at microscopic scale become crucial for fast magnetic reconnection around the diffusion region \citep{birn2001}, making it inherently multiscale. Clarifying the interaction between the MHD-scale physics and the plasma kinetics is one of the most important topics, not only for understanding the physics of fast magnetic reconnection but also for plasma physics as a whole \citep{ji2022, nakamura2025}.
 
Petschek reconnection is one of the representative MHD models of magnetic reconnection \citep{petschek1965}. In this model, magnetic field lines reconnect within a small diffusion region and a pair of switch-off slow shocks are formed outside, converting magnetic energy into plasma energy. The reconnection rate predicted by Petschek model is $\mathcal{O}(0.01 - 0.1)$, which is sufficient to account for the observed timescales of solar flares. Observational studies using such as SDO/AIA \citep{lemen2012}, which provide high-resolution solar images, have shown that similar structures of Petschek reconnection appear during solar flares \citep{masuda1994, tsuneta1996, shiota2003, savage2010, imada2013, warren2018}. Numerical studies have also demonstrated that Petschek reconnection exists as a solution of resistive MHD framework when localized resistivity is introduced \citep{ugai1995, yokoyama2001, shiota2003, zenitani2011petschek, zenitani2015}. However, Petschek reconnection can be realized when the local resistivity is included, and not realized when the uniform resistivity is included \citep{kulsrud2011}. Although the local resistivity is believed to arise from some kinetic effects, whether this assumption is valid remains unclear.

To date, particle-in-cell (PIC) simulations have yet to reproduce the salient features of Petschek reconnection. In particular, no study has reported the existence of the switch-off slow shocks in anti-parallel current sheet composed of ions and electrons \citep{drake2009, liu2012, fujimoto2016, akutagawa2025}. A possible explanation is that the temperature anisotropy of ions at the reconnection exhausts inhibits the formation of switch-off slow shocks, as supported by the analytical theory \citep{hau1993, karimabadi1995, liu2011theory}, simulations of the Riemann problem for modeling the reconnection exhaust \citep{scholer1998, liu2011pic}, and simulations of magnetic reconnection \citep{liu2012, higashimori2012}. It is also found that a weak guide field suppresses temperature anisotropy through pitch angle scattering \citep{le2014}, thereby allowing switch-off slow shocks to develop \citep{innocenti2015}. Moreover, hybrid simulations and observational studies of the Earth magnetosphere have reported the existence of slow shocks, suggesting that Petschek reconnection can be realized in collisionless systems \citep{saito1995, eriksson2004, higashimori2012, walia2022, walia2024}. However, hybrid simulation studies have used localized resistivity, and observational studies have reported only the existence of slow shocks, not switch-off slow shocks. Whether Petschek reconnection exists or not in collisionless systems remains one of the long-standing questions in the magnetic reconnection research.

\revise{}{Previous studies have reported that switch-off slow shocks can form in one-dimensional systems, whereas they do not form in two-dimensional systems \citep{lin1995, lottermoser1998, scholer1998}. This difference is attributed to the presence of free energy in elongated current sheets, which is unstable to the firehose instability. In 2D systems, fluctuations associated with this instability can develop along the current sheet, thereby suppressing the formation of switch-off slow shocks. In contrast, the firehose instability cannot develop in 1D systems, allowing switch-off slow shocks to form easily. Therefore, to understand multidimensional effects, at least 2D simulations are required.} 

In this study, we investigate whether the switch-off slow shocks can be formed in systems where kinetic effects operate locally, using a multi-hierarchy simulation, a method in which the PIC simulation is embedded within the MHD simulation, thereby incorporating plasma kinetics locally \citep{sugiyama2007, usami2013, daldorff2014, makwana2017, haahr2025, akutagawa2025kammuy}. Although this scheme was originally proposed to reduce the computational cost of PIC simulations, the ability to artificially control the region where kinetic effects operate also makes it useful for evaluating how kinetic effects influence the MHD-scale dynamics.

The outline of this paper is as follows. In Section \ref{chap2:simulation_setup}, we explain the simulation setup of multi-hierarchy simulations of two-dimensional Riemann problem which models the outflow region. Section \ref{chap3:results} shows the results and clarifies the dependence of the PIC domain size for the characteristics of the boundary between the inflow and outflow regions. The PIC domain size can be considered to correspond to the mean free path of each system. Section \ref{chap4:conclusion_and_discussion} summarizes the results obtained in this study and discusses their implications.


\section{Simulation setup}\label{chap2:simulation_setup}

The simulation code used in this study is the open-source plasma simulation code KAMMUY \citep{akutagawa2025kammuy}. It includes both MHD and PIC simulation codes, written in CUDA C++ and MPI for multi-GPU acceleration. We perform two-dimensional Riemann problem which models the reconnection outflow region. Riemann problem is often performed because its time evolution mimics the spatial profile of reconnection layer as the distance from X-point increases \citep{lin1995, scholer1998, liu2011pic}.

\subsection{Overview of the Model and Numerical Schemes}\label{chap2:schemes}

Multi-hierarchy simulation treats both MHD and PIC models. The equations of ideal MHD model are as follows: 
\begin{align*}
    \frac{\partial \rho}{\partial t} + \nabla \cdot (\rho \bm{v}) &= 0, \\
    \frac{\partial \rho \bm{v}}{\partial t} + \nabla \cdot \left[ \rho \bm{vv} + \left(p + \frac{|\bm{B}|^2}{2\mu_0} \right) \bm{I} - \frac{\bm{BB}}{\mu_0} \right] &= 0, \\ 
    \frac{\partial \bm{B}}{\partial t} + \nabla \cdot (\bm{vB} - \bm{Bv}) &= 0, \\
    \frac{\partial e}{\partial t} + \nabla \cdot \left[ \left(e + p + \frac{|\bm{B}|^2}{2\mu_0} \right) \bm{v} - \frac{\bm{B}}{\mu_0} \left( \bm{v} \cdot \bm{B} \right) \right] &= 0, 
\end{align*}
where $\rho$ denotes the mass density, $\bm{v} = (v_x, v_y, v_z)$ the bulk velocity, $\bm{B} = (B_x, B_y, B_z)$ the magnetic field, $\bm{E} = (E_x, E_y, E_z)$ the electric field, $\bm{j} = (j_x, j_y, j_z) = \nabla \times \bm{B} / \mu_0$ the current density, $p$ the pressure, $e$ the total energy density, and $\mu_0$ the magnetic permeability of vacuum. The MHD equations are solved using the finite volume method with the HLLD approximate Riemann solver \citep{miyoshi2005}, 2nd-order MUSCL reconstruction \citep{vanleer1979}, and a 2nd-order Runge-Kutta method. The solenoidal condition of the magnetic field is enforced using the projection method \citep{brackbill1980}.

The equations of PIC model are as follows: 
\begin{align*}
    \frac{d \bm{x}_{\rm s}}{dt} &= \bm{v}_{\rm s}, \\
    \frac{d}{dt} (\gamma m_{\rm s} \bm{v}_{\rm s}) &= q_{\rm s} \left( \bm{E} (\bm{x}_{\rm s}) + \bm{v}_{\rm s} \times \bm{B} (\bm{x}_{\rm s}) \right), \\
    \frac{\partial \bm{B}}{\partial t} &= -\nabla \times \bm{E}, \\
    \frac{\partial \bm{E}}{\partial t} &= c^2 \nabla \times \bm{B} - \frac{\bm{j}}{\epsilon_0} + d \nabla F, \\
    \bm{j} &= \sum_s q_s \int f_s \bm{v}^\prime d\bm{v}^\prime, \\
    F &= \nabla \cdot \bm{E} - \rho_q, \\
    \rho_q &= \sum_s q_s \int f_s d\bm{v}^\prime, 
\end{align*}
where $\bm{x}_s$ and $\bm{v}_s$ denote the position and velocity, $\gamma$ the Lorentz factor of the particle, $m_s$ the rest mass, $q_s$ the charge of particle species $s$, $\rho_q$ the charge density, $c$ the light speed, and $\epsilon_0$ the dielectric constant. The equations of motion are solved using the relativistically extended Buneman–Boris method \citep{boris1970}, which ensures second-order accuracy in time and numerical stability except for ultra-relativistic particles. Maxwell's equations are solved with the Yee lattice \citep{yee1966} and the Leapfrog scheme, which are second-order accurate in space and time. Charge and current densities are obtained from the distribution function using linear interpolation (CIC) following \citet{birdsall_and_langdon_1991}. We also apply the Langdon–Marder correction \citep{marder1987, langdon1992} to eliminate Poisson noise; the error of Poisson equation $F$ is introduced and a new term is included on the right hand side of the Amp\`ere's law.

Multi-hierarchy simulation treats MHD and PIC simultaneously using the interlocking method. Physical quantities are exchanged between MHD and PIC. PIC domain has interface region around the boundary, and physical quantities of MHD are sent to PIC at the interface region. Physical quantities of PIC are averaged and sent to MHD domain where PIC is embedded. In this paper, we use $3 \times 3$ averaging filter for MHD domain once per 10 MHD steps in all simulations to remove artificial noise from the projection method. The details and validations of multi-hierarchy scheme are written in \cite{akutagawa2025kammuy}. 

\subsection{Initial Condition}\label{chap2:initial_condition}

We use a two-dimensional force-free current sheet with a finite $B_y$ component. In this paper, we set $B_y = 0.1 B_0$ where $B_0$ is the strength of the anti-parallel magnetic field. It comes from previous studies that the reconnection rate for collisionless magnetic reconnection is about $0.1$ \citep{birn2001, zenitani2011pop}. Specifically, each component of the initial magnetic field is given as follows.
\begin{align*}
    B_x &= B_{\rm 0} \tanh (y / \delta), \\
    B_y &= 0.1 B_{\rm 0}, \\
    B_z &= B_{\rm 0} / \cosh (y / \delta),
\end{align*}    
where $\delta$ is the half-thickness of the current sheet, and plasmas are uniformly distributed. The half-thickness is set to $\delta = 2.0 \lambda_{\rm i}$ where $\lambda_{\rm i}$ is the ion inertial length. The upstream plasma beta is set to $\beta = 0.25$.

We set $m_{\rm i} / m_{\rm e} = 25$, $n_{\rm i0} = n_{\rm e0} = 20 \ \text{ppc}$, $T_{\rm i0}/T_{\rm e0} = 1$, and $\omega_{\rm pe} / \Omega_{\rm ce} = 1$. 
Previous study reported that the reconnection rate for two-dimensional collisionless magnetic reconnection is independent of $m_{\rm i} / m_{\rm e}$ and $\omega_{\rm pe} / \Omega_{\rm ce}$. 
The size of the PIC simulation domain is $2000 \times 100, 200, 400$, and $800$ for the PIC domain. Grid-size ratio of MHD and PIC is adopted as $\Delta_{\rm MHD} / \Delta_{\rm PIC} = 10$ in whole simulations. The size of the MHD domain is $200 \times 1000$ (corresponding to $2000 \times 10000$ PIC grids). It corresponds to $100\lambda_{\rm i} \times 500\lambda_{\rm i}$ simulation box. The periodic boundary condition is imposed in the $x$ direction, while the symmetric boundary condition is imposed in the $y$ direction.

We note that, in our multi-hierarchy model with $N_{y, {\rm PIC}} = 100$, the total number of particles required to cover the simulation domain is reduced to about $1/100$ of that required in a full-PIC simulation The most computationally expensive component of PIC simulations is the calculation of moments, and thus the total number of particles determines the overall computational cost. Consequently, the total computational cost is reduced to about $1/100$ of that of a full-PIC simulation, allowing us to carry out large-domain simulations that are impractical for full-PIC models.


\section{Results}\label{chap3:results}

\subsection{Overview of Current Sheet Structure}\label{chap3:overview}

\begin{figure}[ht!]
    \centering
    \includegraphics[width=\linewidth]{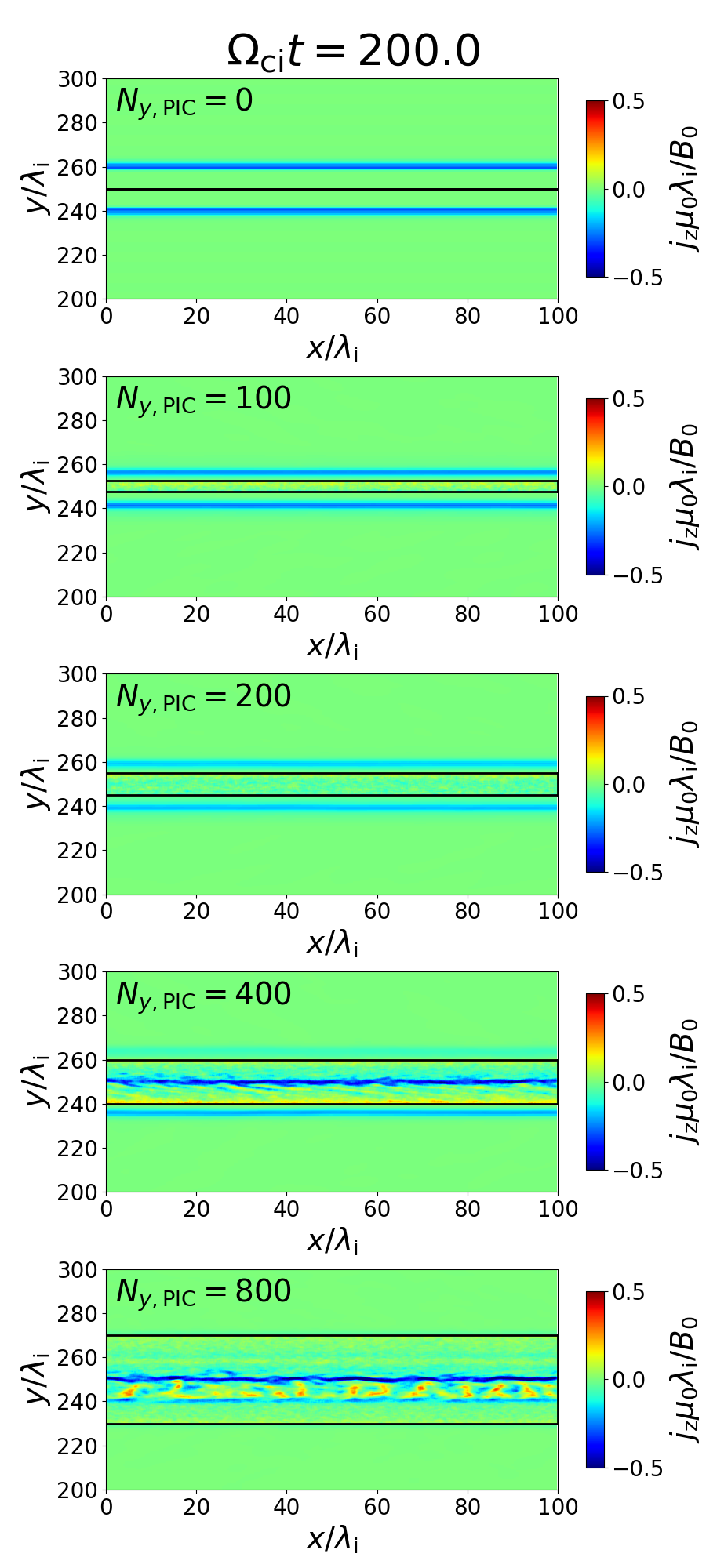}
    \caption{Spatial profiles of $j_{z}$ from ideal MHD and multi-hierarchy simulations at $\Omega_{\rm ci} t = 200.0$. PIC data is used between the black lines, and MHD data is used outside. $j_{z}$ is normalized by $B_{\rm 0} / \mu_{\rm 0} \lambda_{\rm i}$. The horizontal and vertical axes are normalized by $\lambda_{\rm i}$.}
    \label{chap3:fig:overview_jz}
\end{figure}

Figure \ref{chap3:fig:overview_jz} shows the spatial profiles of $j_{z}$ at $\Omega_{\rm ci} t = 200.0$ from ideal MHD ($N_{y, \rm PIC} = 0$) and four multi-hierarchy simulations ($N_{y, {\rm PIC}} = 100, 200, 400, 800$). In ideal MHD simulation, we find two bifurcated current sheets, corresponding to the switch-off slow shocks in the conventional Petschek reconnection. A similar structure is also found in the multi-hierarchy simulations with $N_{y, {\rm PIC}} = 100$ and $200$.

In contrast, for $N_{y, {\rm PIC}} = 400$ and $800$, the current sheet at the center is dominant, corresponding to the elongated current sheet commonly reported in full PIC simulation of magnetic reconnection \citep{liu2012, le2014, fujimoto2016}. \cite{le2016} reported that temperature anisotropy is important for the elongated current sheet formation, using equations of state for isotropic and anisotropic plasmas. These results indicate that anisotropic plasmas exist for $N_{y, {\rm PIC}} = 400$ and $800$ at $\Omega_{\rm ci} t = 200.0$. We will discuss temperature anisotropy in subsection \ref{chap3:isotropization}.


\subsection{Spatio-temporal diagrams of $j_z$ and $\nabla \cdot \boldsymbol{v}$}\label{chap3:spatio-temporal_diagram_of_jz_and_divV}

\begin{figure*}[ht!]
    \centering
    \includegraphics[width=\linewidth]{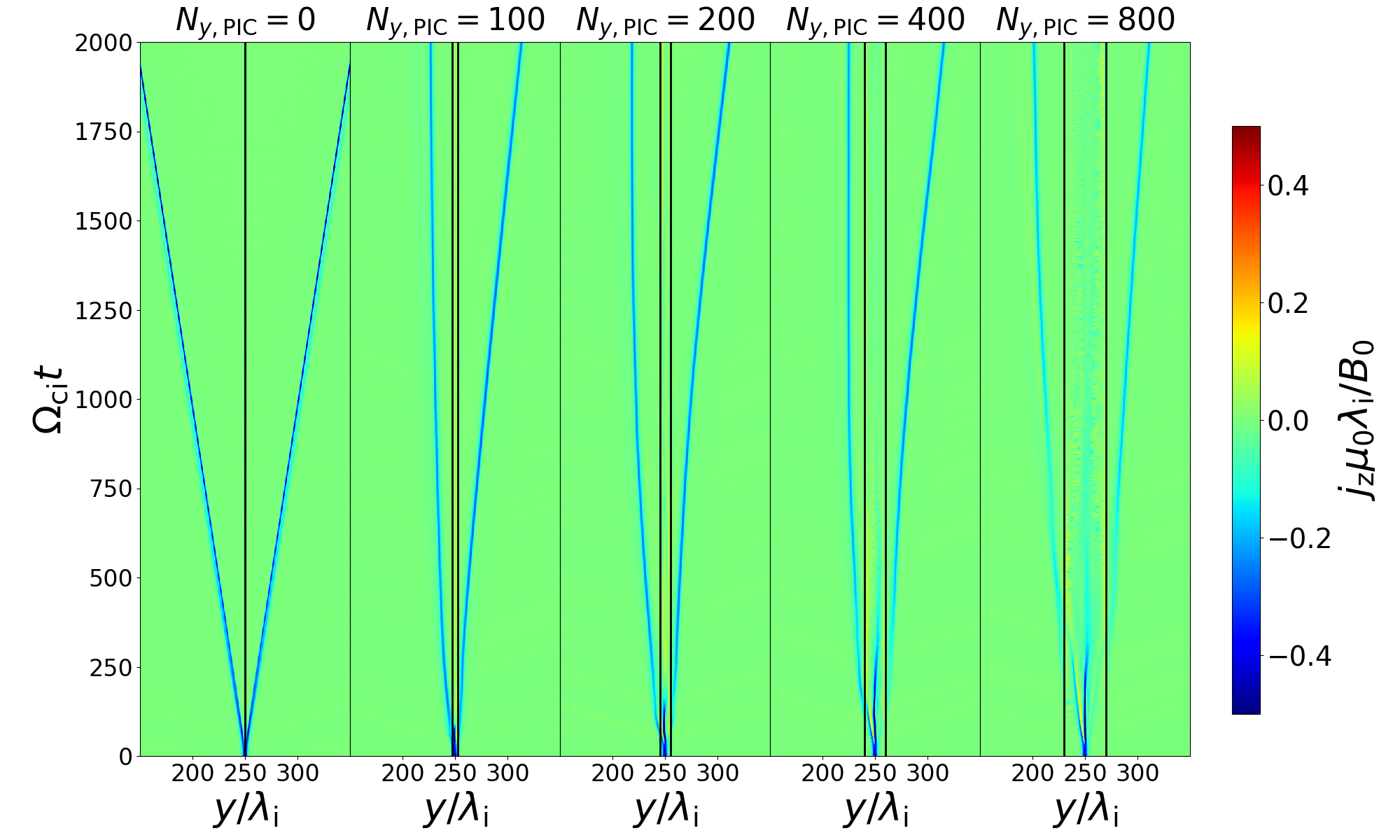}
    \includegraphics[width=\linewidth]{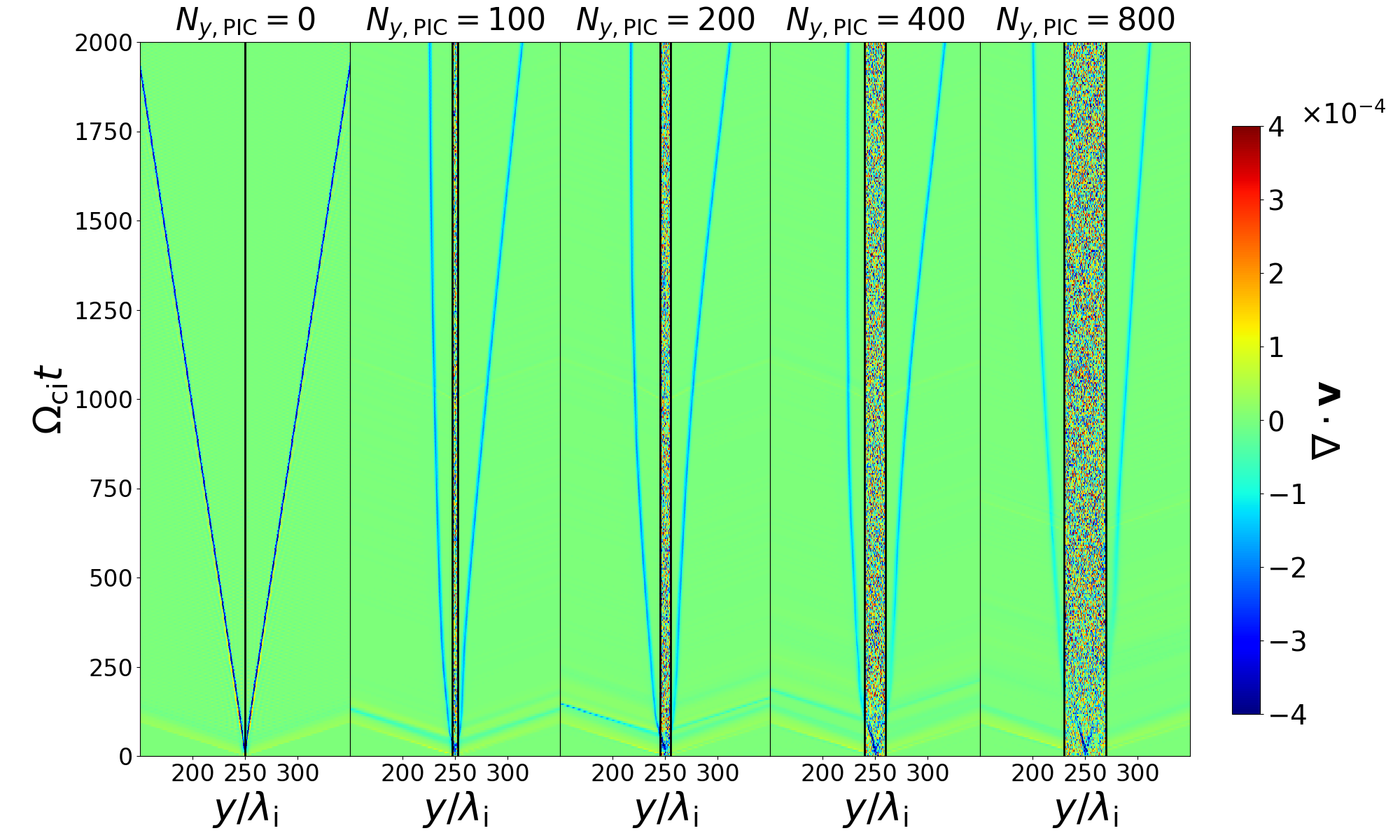}
    \caption{$y - t$ diagrams of $j_{z}$ and $\nabla \cdot \bm{v}$ averaged in the $x$ direction. The panels show the results for $N_{y, \rm PIC} = 0$ (ideal MHD), $100, 200, 400$, and $800$. $j_{z}$ is normalized by $B_{\rm 0} / \mu_{\rm 0} \lambda_{\rm i}$. $\bm{v}$ is calculated using the zeroth and first moment of ions and electrons. The horizontal and vertical axes are normalized by $\lambda_{\rm i}$ and $\Omega_{\rm ci}^{-1}$ respectively.}
    \label{chap3:fig:xt_diagram_jz_and_divV}
\end{figure*}

The upper panel of Figure \ref{chap3:fig:xt_diagram_jz_and_divV} shows the $y - t$ diagram of $j_z$ averaged in the $x$ direction. The result of the ideal MHD simulation (where $N_{y, {\rm PIC}} = 0$) shows that the bifurcated current sheets, corresponding to the switch-off slow shocks, extend straight outward from the center. This behavior reflects the hyperbolic nature of the MHD equations. Similar structures are clearly seen in the multi-hierarchy simulation result for $N_{y, {\rm PIC}} = 100$; however, the slope is different and shows a pronounced asymmetry. We discuss this asymmetry in the Appendix \ref{appendix:asymmetry_of_the_y-t_diagram}. At $\Omega_{\rm ci} t \sim 500$, as the PIC domain size $N_{y, {\rm PIC}}$ increases from $100$ to $800$, the peak value of $j_z$ is reduced. These results indicate that kinetic effects work to suppress the expansion of the current sheet in the early phase. However, the peak value becomes larger at later time $\Omega_{\rm ci} t \gtrsim 1500$ for $N_{y, {\rm PIC}} = 800$. It suggests that the $j_z$ peak can be appeared even when the PIC domain is sufficiently large.

The lower panel of Figure \ref{chap3:fig:xt_diagram_jz_and_divV} shows the $y - t$ diagram of $\nabla \cdot \bm{v}$ averaged in the $x$ direction. At $\Omega_{\rm ci} t \sim 2000$ for all $N_{y, \rm PIC}$, regions with $\nabla \cdot \bm{v} < 0$ develop in the MHD domain and appear at the same locations as the current sheet. These results indicate that the current sheets are produced by MHD shocks given sufficient time, and the formation of the shock does not depend on the PIC domain size. Moreover, the propagation speed of $\nabla \cdot \bm{v}$ is $\lesssim 100\lambda_{\rm i} / 2000 \Omega_{\rm ci}^{-1} \ll V_{\rm A0}$. This suggests that slow shocks are formed in all multi-hierarchy simulation results.


\subsection{Properties of slow shock}\label{chap3:properties_of_slow_shock}

In subsection \ref{chap3:spatio-temporal_diagram_of_jz_and_divV}, it is found that the boundary between the inflow and outflow region becomes slow shock in the MHD domain for any PIC domain size. However, whether the slow shock is switch-off slow shock or not is unknown. We analyze the properties of the slow shocks formed in the MHD domain by comparing the switch-off slow shocks formed in the ideal MHD simulation result and using Rankine-Hugoniot (RH) relation in anisotropic plasmas.

\begin{figure*}[ht!]
    \centering
    \includegraphics[width=0.45\linewidth]{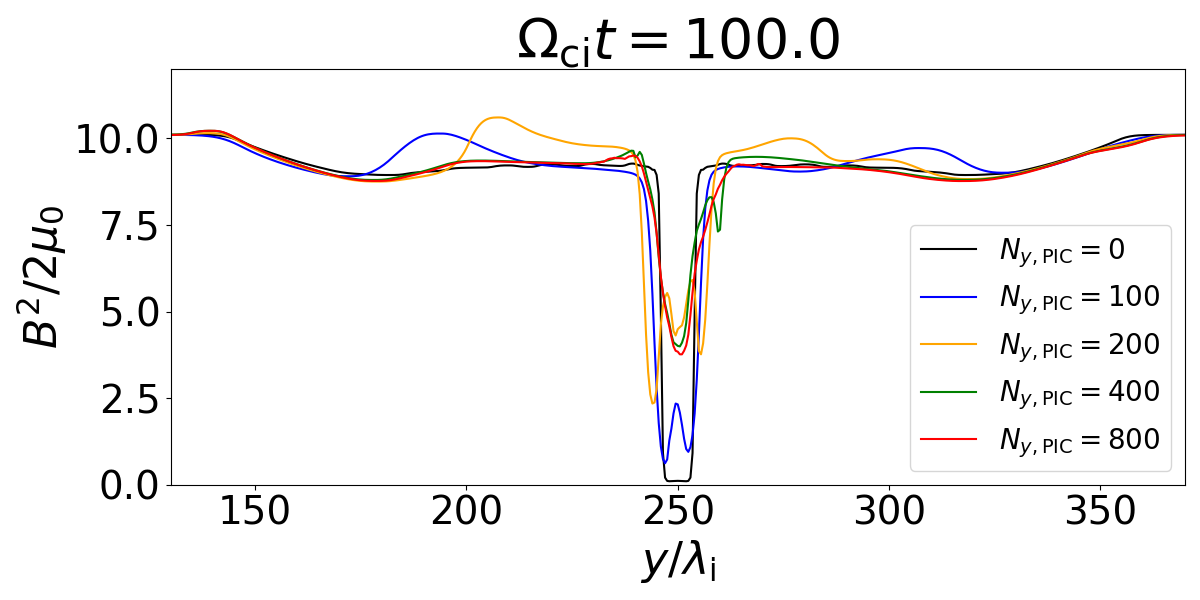}
    \includegraphics[width=0.45\linewidth]{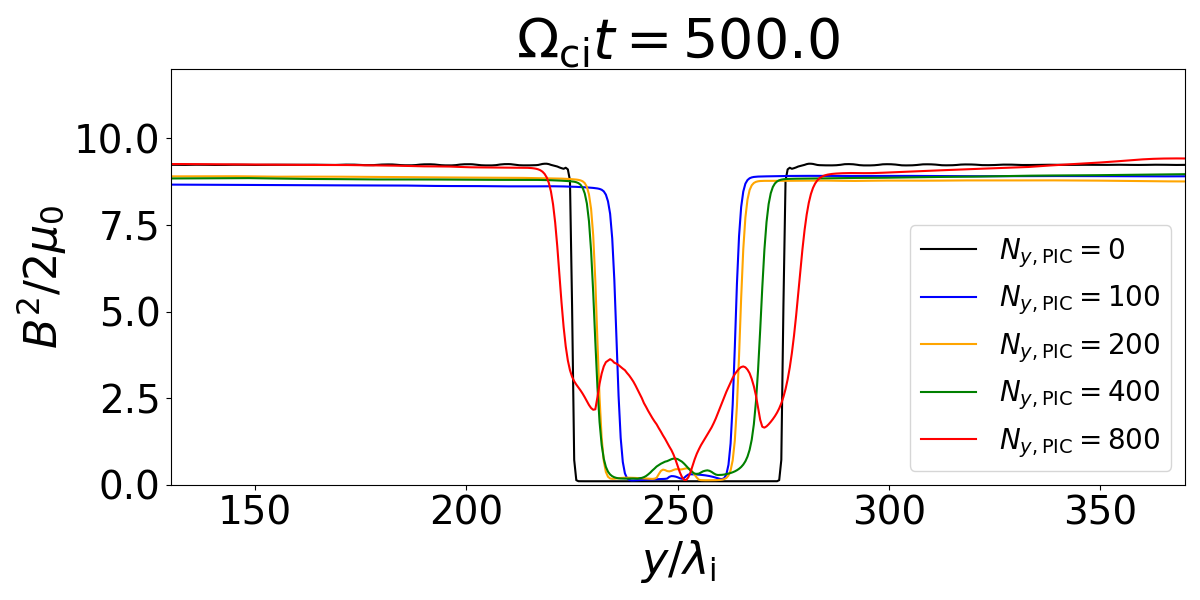}
    \includegraphics[width=0.45\linewidth]{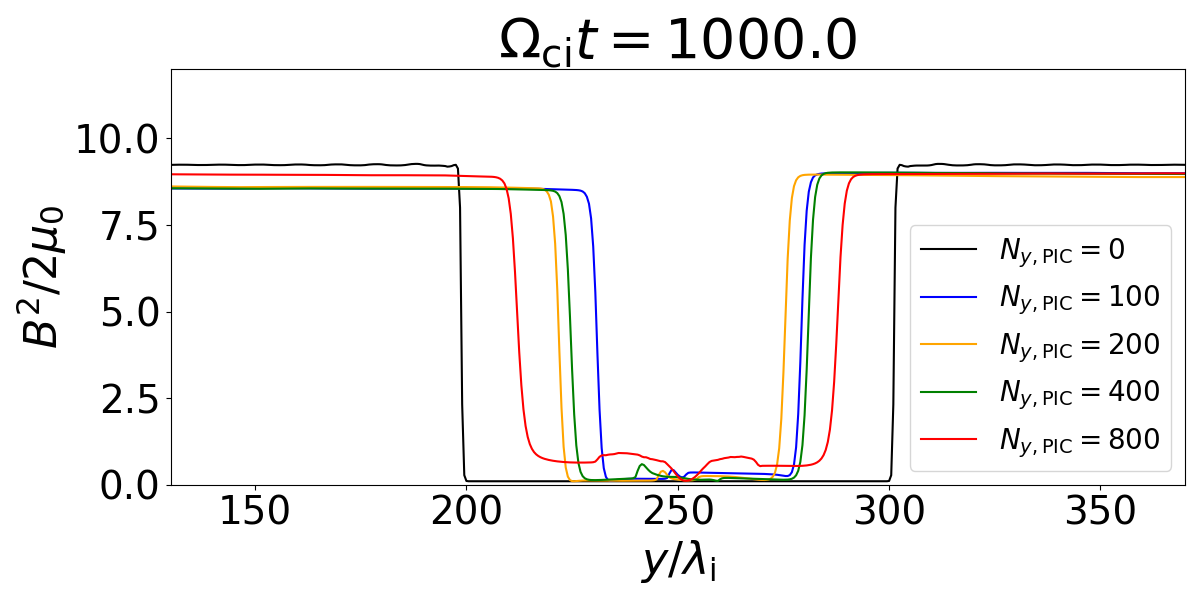}
    \includegraphics[width=0.45\linewidth]{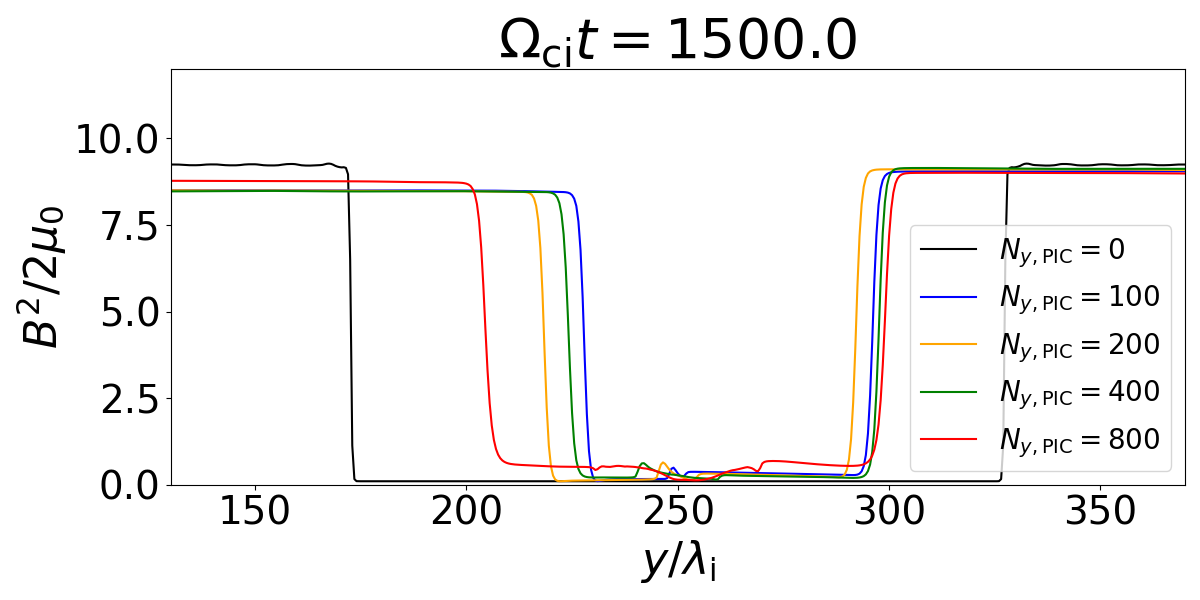}
    \caption{Snapshots of magnetic energy $B^2 / 2 \mu_{\rm 0}$ from ideal MHD and multi-hierarchy simulations at $\Omega_{\rm ci} t = 100.0, 500.0, 1000.0$, and $1500.0$. The magnetic field from the MHD data is only used, and averaged in the x direction. The horizontal axis is normalized by $\lambda_{\rm i}$.}
    \label{chap3:fig:slice_diagram_magnetic_energy}
\end{figure*}

Figure \ref{chap3:fig:slice_diagram_magnetic_energy} shows magnetic energy averaged in the $x$ direction at $\Omega_{\rm ci} t = 100.0, 500.0, 1000.0$, and $1500.0$. At $\Omega_{\rm ci} t = 100.0$, fast mode rarefaction waves propagate outward ($y \sim 150 \lambda_{\rm i}$ and $350 \lambda_{\rm i}$) and reduce magnetic energy slightly. Although magnetic energy is decreasing around the center ($y \sim 240 \lambda_{\rm i}$ and $260 \lambda_{\rm i}$), the reduction is small compared to the ideal MHD simulation result (black line) corresponding to the switch-off slow shock. At the relatively late stage of $\Omega_{\rm ci} t = 1500.0$, the magnetic energy decreasing becomes similar to that observed in the ideal MHD simulation. Similar reduction of magnetic energy in any PIC domain size suggests that the switch-off slow shocks may form in the MHD domain and be independent of the PIC domain size.

\begin{figure}[ht!]
    \centering
    \includegraphics[width=0.9\linewidth]{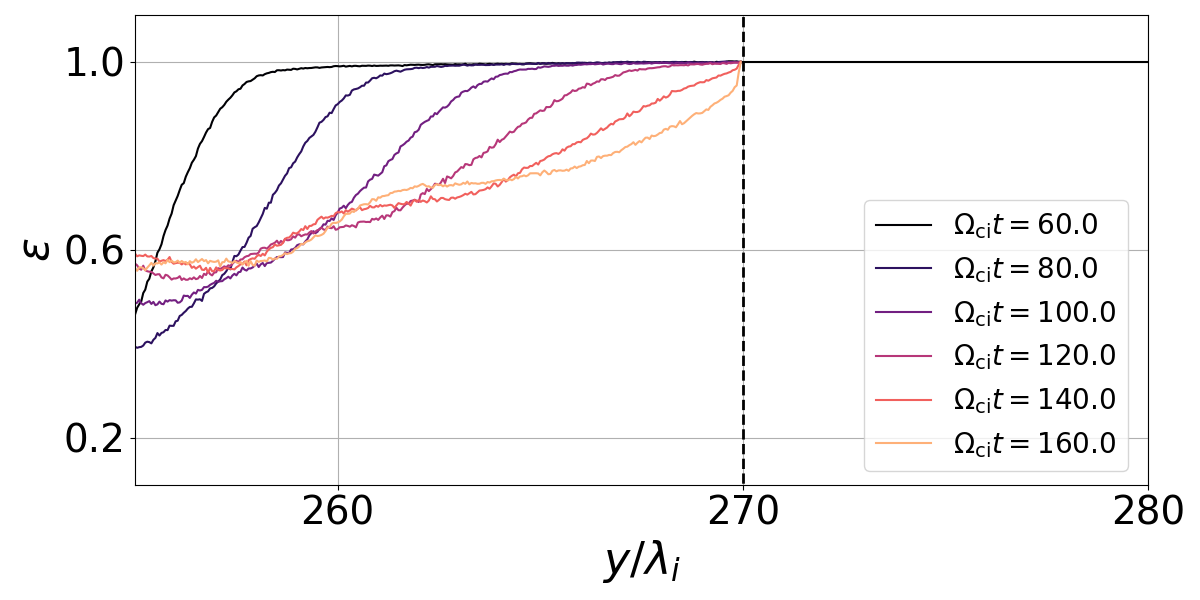}
    \includegraphics[width=0.9\linewidth]{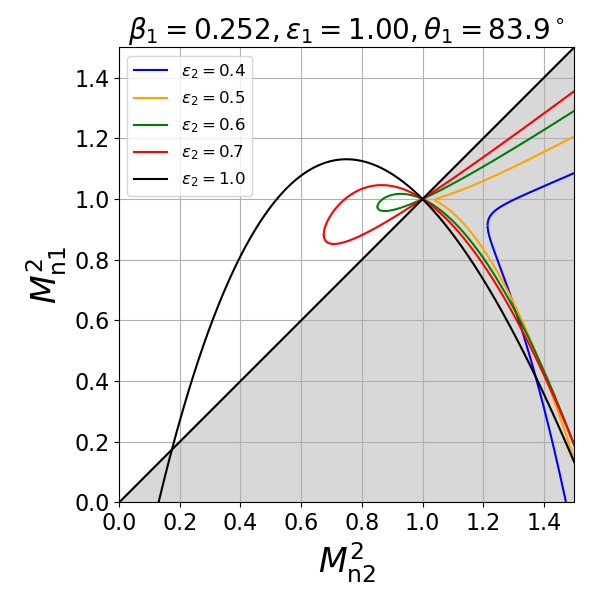}
    \includegraphics[width=0.9\linewidth]{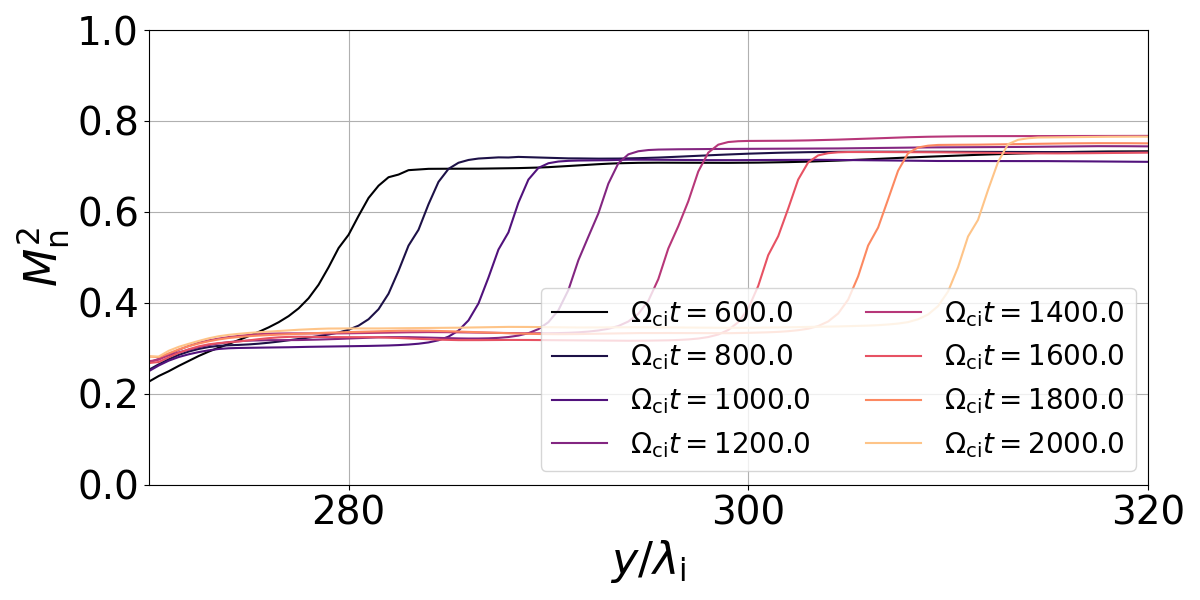}
    \caption{Top panel: Six snapshots of the firehose stability parameter $\epsilon$ from the multi-hierarchy simulation with $N_{y, \rm PIC} = 800$. The horizontal axis is normalized by $\lambda_{\rm i}$. Middle panel: Upstream--downstream Mach number difference $M_{\rm n1}^2 - M_{\rm n2}^2$ derived from the Rankine--Hugoniot relations for anisotropic plasmas. Subscripts 1 and 2 denote the upstream and downstream states, respectively. Bottom panel: Eight snapshots of the $M_{\rm n}$ in the MHD domain for the same simulation. The horizontal axis is normalized by $\lambda_{\rm i}$.}
    \label{chap3:fig:epsilon_RH_Mn}
\end{figure}

To understand why the structure like switch-off slow shock is formed, we analyze the RH relation in anisotropic plasmas to investigate the quantitative conditions for the slow shock solution. The upper panel of Figure \ref{chap3:fig:epsilon_RH_Mn} shows the time evolution of the firehose stability parameter $\epsilon := 1 - \mu_{\rm 0} (P_\parallel - P_\perp) / B^2$ averaged in the $x$ direction. The region of decreasing $\epsilon$ indicates the boundary between the inflow and outflow region, which is seen in previous studies \citep{liu2011pic, liu2012, higashimori2012}. The reason is written in subsection \ref{chap3:isotropization}. $\epsilon \sim 0.5$ at the downstream region when the boundary begins to enter the MHD domain ($\Omega_{\mathrm{ci}} t = 300 - 400$ ). After the boundary entering the MHD domain, $\epsilon$ increases.

The middle panel of Figure \ref{chap3:fig:epsilon_RH_Mn} shows the solutions of $M_{n1}^2 - M_{n2}^2$ from anisotropic MHD equations \citep{karimabadi1995, higashimori2012}. Here, $M_{n} := v_{\rm n} / \sqrt{\epsilon} V_{\rm An}$ and $V_{\rm An} := B_n / \sqrt{\mu_0 \rho}$. $Q_{\rm n}$ denotes the normal component of the physical quantity $Q$. The subscripts $1$ and $2$ denote the upstream and downstream respectively. $\beta_1$, $\epsilon_1$ and $\theta_1$ are set from the upstream data after the propagation of the fast mode rarefaction wave; $\beta_1 = 0.252$, $\epsilon_1 = 1.00$, and $\theta_1 = 83.9^\circ$. Details of how to derive $M_{\rm n1}^2 - M_{\rm n2}^2$ relation is written in the Appendix \ref{appendix:anisotropic_plasmas_RH}. When the boundary between the inflow and outflow region is inside the PIC domain, $\epsilon_2$ is nearly $0.6$. The conditions for satisfying the anisotropic RH relation become quite stringent. 

The lower panel of Figure \ref{chap3:fig:epsilon_RH_Mn} shows $M_{\rm n}^2$ calculated from the MHD data after the boundary entering the MHD domain. In calculating the shock propagating speed $V_{\rm sh}$, we use the continuity equation of mass density as follows:
\begin{align*}
    V_{\rm sh} = \frac{\rho_2 v_{y2} - \rho_1 v_{y1}}{\rho_2 - \rho_1}.  
\end{align*}
We average the data in the $x$ direction and use $\rho$ and $v_y$ at $y = 275 \lambda_{\rm i}$ (downstream) and $y = 320 \lambda_{\rm i}$ (upstream). 
The downstream plasma becomes isotropic so that the property of the shock lies on the solution of $\epsilon_2 = 1.0$. The upstream and downstream Mach number squared are $M_{\rm n1}^2 \sim 0.7$ and $M_{\rm n2}^2 \sim 0.3$, and the property of the boundary lies on the slow shock solution of $\epsilon_2 = 1.0$. The absence of an anisotropic RH solution satisfying $M_{\rm n1}^2 \sim 0.7$ at $\epsilon_2 \sim 0.6$ indicates that the boundary between the inflow and outflow regions is not a shock, but a pulse wave. We discuss the property of this pulse wave in subsction \ref{chap3:mode_conversion_of_pulse_wave}.


\subsection{Isotropization}\label{chap3:isotropization}

In subsections~\ref{chap3:spatio-temporal_diagram_of_jz_and_divV} and \ref{chap3:properties_of_slow_shock}, we show that the boundary between the inflow and outflow regions develops a structure consistent with a switch-off slow shock. This structure is insensitive to the size of the PIC domain. These findings indicate that Petschek-like reconnection can be sustained even when kinetic effects are included locally. The above results primarily address the influence of the PIC domain on the surrounding MHD domain. By contrast, the reverse influence, namely how the large-scale MHD environment affects the plasma properties within the PIC domain, remains unclear, particularly with regard to the role of slow-shock formation. In this subsection, we analyze this reverse coupling in detail. 

\begin{figure}[t!]
    \centering
    \includegraphics[width=\linewidth]{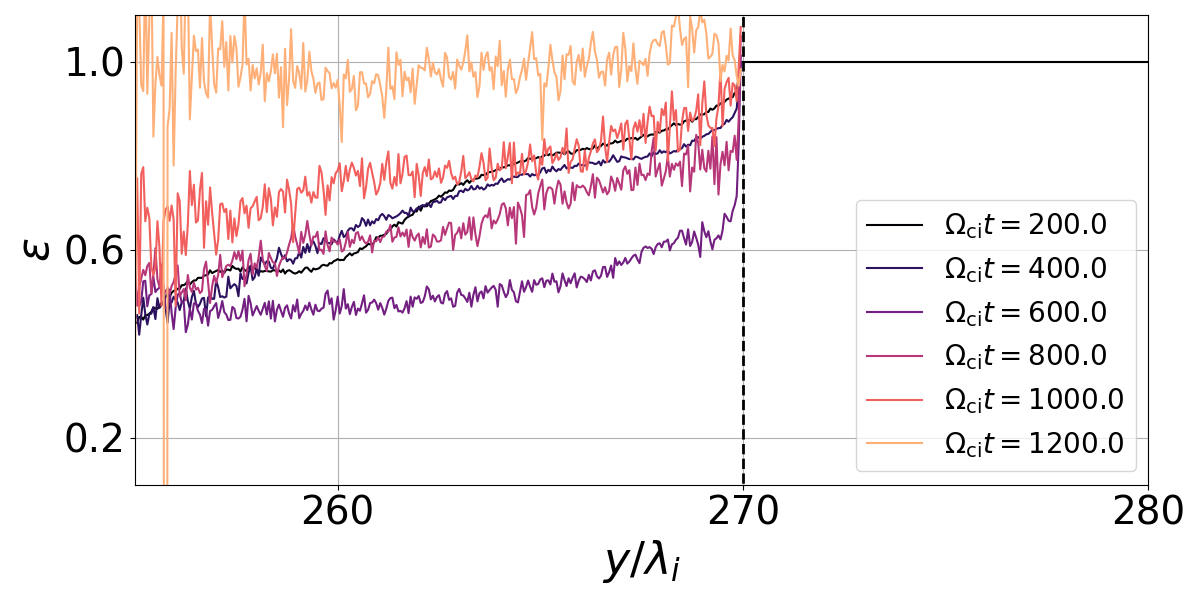}
    \includegraphics[width=\linewidth]{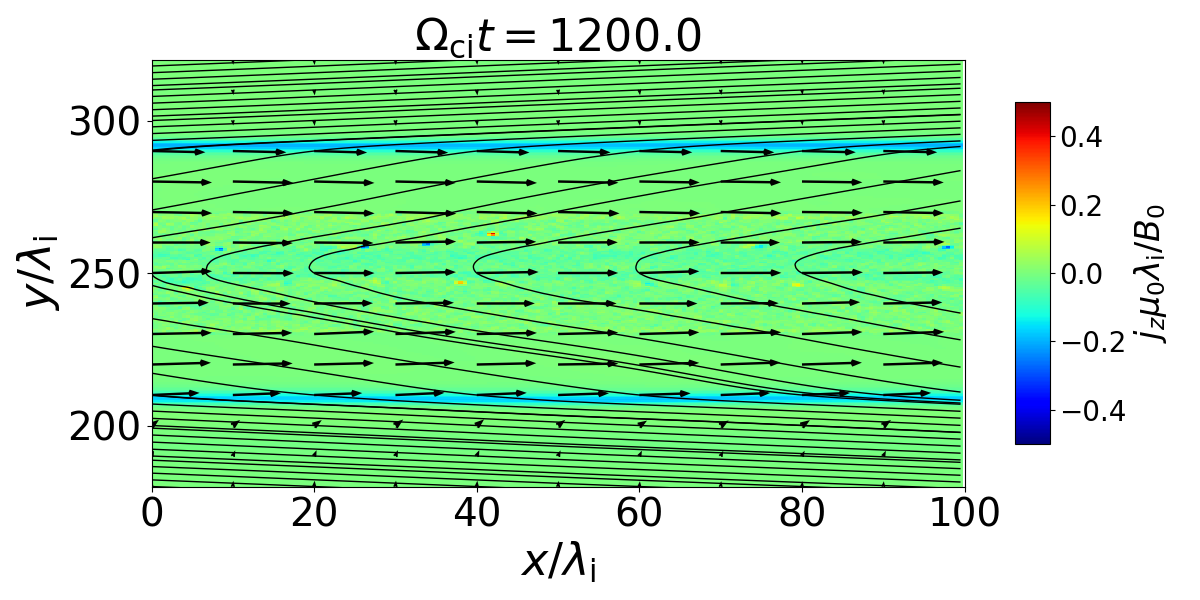}
    \caption{
    Top panel: Six snapshots of the firehose stability parameter $\epsilon$ in the multi-hierarchy simulation with $N_{y, \rm PIC} = 800$. The horizontal axis normalized by $\lambda_{\rm i}$. Bottom panel: Spatial profile of $j_z$ at $\Omega_{\rm ci} t = 1200.0$ for the same simulation. The black solid curve traces the magnetic field line, and the black arrow indicates the bulk velocity. The current density $j_z$ is normalized by $B_0 / \mu_0 \lambda_{\rm i}$. Horizontal and vertical axes are normalized by $\lambda_{\rm i}$.}
    \label{chap3:fig:epsilon_jz}
\end{figure}

The upper panel of Figure \ref{chap3:fig:epsilon_jz} shows the time evolution of $\epsilon$ for $N_{y,\rm PIC} = 800$. In the early stage, $\epsilon < 1$, whereas it gradually approaches unity at later times ($\Omega_{\rm ci} t \gtrsim 1200$). As shown in the upper panel of Figure \ref{chap3:fig:xt_diagram_jz_and_divV}, the central peak in $j_z$ begins to weaken at $\Omega_{\rm ci} t \sim 500$, which coincides with the onset of the increase in $\epsilon$ toward unity. The lower panel of Figure \ref{chap3:fig:epsilon_jz} presents the spatial profiles of $j_z$, the magnetic field, and the bulk velocity. Across the slow shocks, the magnetic field strength decreases. The curvature of the magnetic field lines near the center is reduced, and the central $j_z$ peak present in the initial stage disappears (see the lowest panel of Figure \ref{chap3:fig:overview_jz}). In addition, the $v_x$ component becomes dominant over $v_y$. These results suggest a causal connection among the decrease in $v_y$, the increase in $v_x$, the reductions in magnetic field curvature and $j_z$ in the outflow region, and the approach of $\epsilon$ toward unity. 

A causal connection among these processes has been suggested in previous numerical and observational studies. Finite ion $v_y$ arising near the boundaries between the inflow and outflow regions can produce multiple ion orbits and generate temperature anisotropy \citep{hoshino1998, gosling2005}. In addition, \cite{le2016} reported that an anisotropic equation of state is required to form an elongated current sheet, whereas such a structure does not develop when an isotropic equation of state is employed. Our results are consistent with these studies. When $\epsilon$ approaches unity, the $v_x$ component dominates in the outflow region and the elongated current sheet disappears.

\subsection{Mode conversion of pulse wave}\label{chap3:mode_conversion_of_pulse_wave}

To understand the property of the pulse wave formed between the inflow and outflow regions in the PIc domain, we perform linear analysis in anisotropic plasma model. Using Chew-Goldberger-Low \citep{chew1956} closure, \cite{hau1993} derived the dispersion relation and the group speeds of fast, slow, and intermediate modes in anisotropic plasmas as follows:
\begin{align*}
    V_{\rm F,S} &= \sqrt{\frac{1}{2} (b \pm \sqrt{b^2 - 4 c})}, \\ 
    b &= V_{\rm A}^2 + C_{\rm S, \perp}^2 + \left [\frac{\gamma_\parallel - 1}{\gamma_\parallel} C_{\rm S, \parallel}^2 - \frac{\gamma_\perp - 1}{\gamma_\perp} C_{\rm S, \perp}^2 \right] \cos^2 \theta, \\
    c &= -\left [C_{\rm S, \parallel}^2 (C_{\rm S, \parallel}^2 \cos^2 \theta - b) + \frac{C_{\rm S, \perp}^4 \sin^2 \theta}{\gamma_\perp^2} \right] \cos^2 \theta, \\
    V_{\rm I} &= \sqrt{V_{\rm A}^2 \epsilon \cos^2 \theta}, \\ 
\end{align*}
where $V_A := B / \sqrt{\mu_0 \rho}$, $C_{\rm S, \perp} := \sqrt{\gamma_\perp p_\perp / \rho}$, and $C_{\rm S, \parallel} := \sqrt{\gamma_\parallel p_\parallel / \rho}$. $\theta$ is the angle between the background magnetic field and wavenumber vector. The degrees of freedom in the perpendicular and parallel directions are $f_\perp = 2$ and $f_\parallel = 1$, respectively, so that $\gamma_\perp = (f_\perp + 2) / f_\perp = 2.0$ and $\gamma_\parallel = (f_\parallel + 2) / f_\parallel = 3.0$. 
For isotropic plasmas, 
\begin{align*}
    V_{\rm F, S} &= \sqrt{\frac{1}{2} (V_{\rm A}^2 + C_{\rm S}^2) \pm \sqrt{\frac{1}{2} [(V_{\rm A}^2 - C_{\rm S}^2)^2 + 4 V_{\rm A}^2 C_{\rm S}^2 \sin^2 \theta]}}, \\ 
    V_{\rm I} &= \sqrt{V_{\rm A}^2 \cos^2 \theta}, 
\end{align*}
where $C_{\rm S} := \sqrt{\gamma p / \rho}$. We use $\gamma = 5.0 / 3.0$.

\begin{figure}[ht!]
    \centering
    \includegraphics[width=\linewidth]{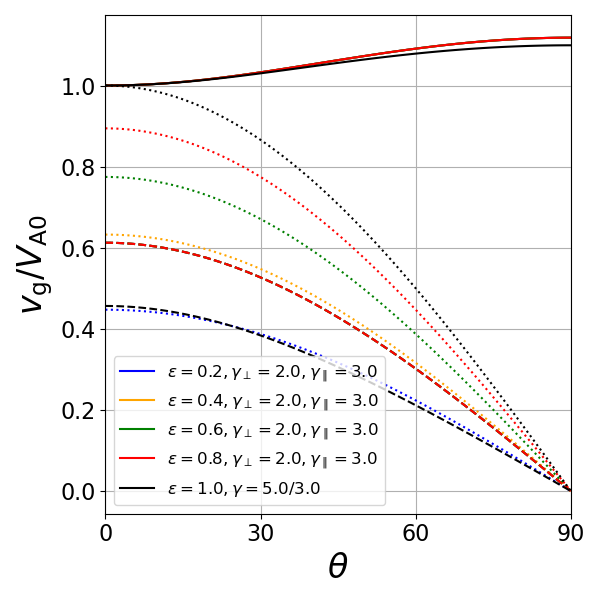}
    \caption{$v_g - \theta$ diagram from anisotropic plasmas with different $\epsilon$ and isotropic plasmas. Dashed, dotted, and solid lines correspond to the slow, intermediate, and fast mode.}
    \label{chap3:fig:group_velocity}
\end{figure}

Figure \ref{chap3:fig:group_velocity} shows the group velocity of anisotropic and isotropic low $\beta$ plasmas as a function of $\theta$. In anisotropic plasmas, the intermediate mode speed approaches the slow mode speed and they degenerate near $\epsilon \sim 0.4$. Furthermore, as the anisotropy increases, their speeds reverse. Figure \ref{chap3:fig:epsilon_RH_Mn} shows that the pulse wave propagates in anisotropic plasmas with large $\theta$, indicating that the pulse wave is a compound wave of slow and intermediate wave. This has already been pointed out theoretically from previous studies using different models \citep{liu2011theory}.

\begin{figure}[ht!]
    \centering
    \includegraphics[width=\linewidth]{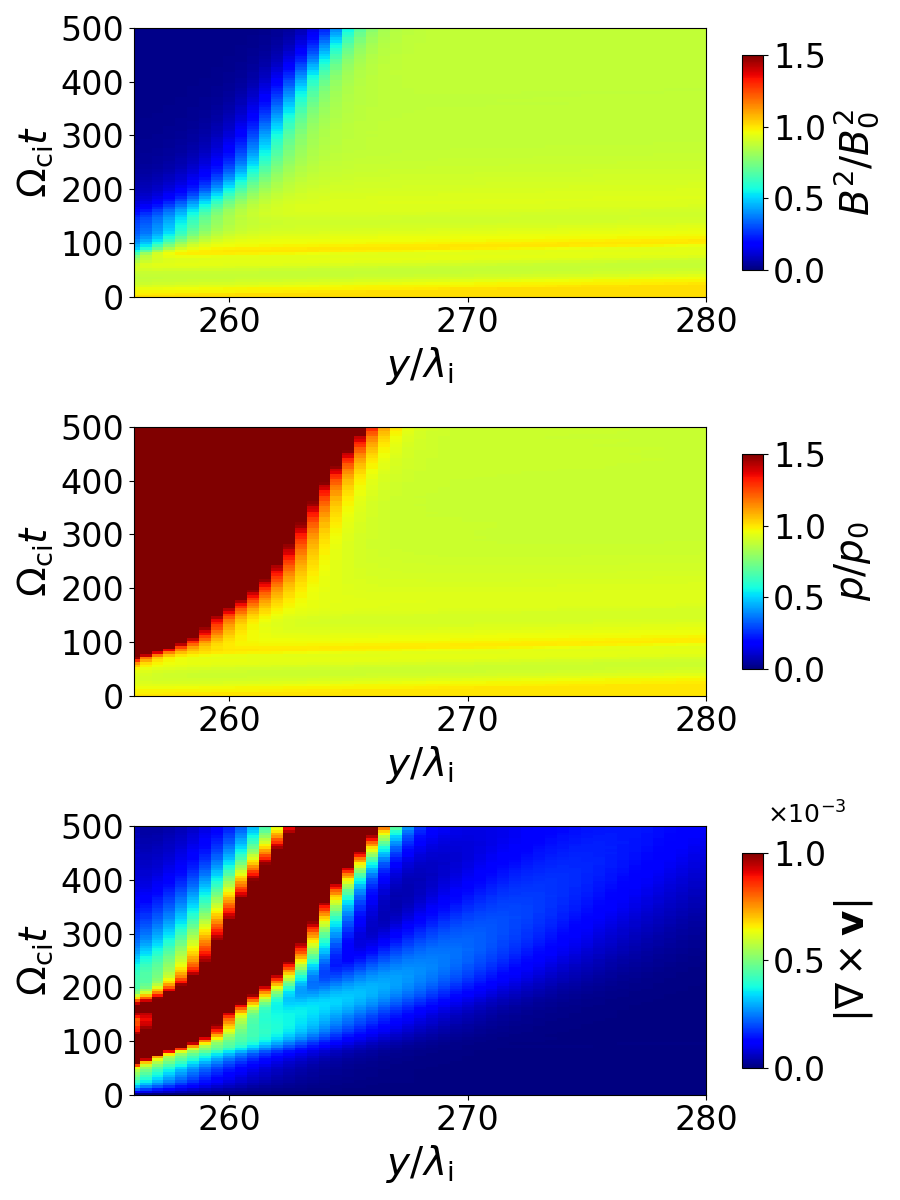}
    \caption{$y - t$ diagrams of magnetic pressure, gas pressure, and vorticity averaged in the $x$ direction. The panels show the result for $N_{y, \rm PIC} = 200$. The left boundary ($y = 255 \lambda_{\rm i}$) is just outside the PIC domain. The horizontal and vertical axes are normalized by $\lambda_{\rm i}$ and $\Omega_{\rm ci}^{-1}$ respectively.}
    \label{chap3:fig:xt_diagram_mode_detection}
\end{figure}

Figure \ref{chap3:fig:xt_diagram_mode_detection} shows the $y - t$ diagram of magnetic pressure, gas pressure, and vorticity averaged in the $x$ direction. The pulse wave enters the MHD domain at $\Omega_{\rm ci} t \sim 50$. After that, two kinds of waves are identified; one is a high-speed propagating wave, for which the slope in the $y - t$ diagram is similar to that of a fast mode rarefaction wave seen at $\Omega_{\rm ci} t \lesssim 50$, and second is a slow-speed propagating wave, for which the amplitude is much larger than that of the former wave. From the behaviors of magnetic and gas pressure fluctuations, the former is the fast mode wave and the latter is the slow mode wave. The $y - t$ diagram of vorticity shows different peak between the slow and fast mode wave, indicating the existence of an intermediate mode wave in the MHD domain. From these results, we conclude that the slow and intermediate compound pulse wave is converted into the large-amplitude slow mode wave, and small-amplitude intermediate and fast mode waves. Large amplitude slow mode wave can steepen to evolve into a slow shock.


\section{Conclusion and Discussion}\label{chap4:conclusion_and_discussion}

\begin{figure*}[ht!]
    \centering
    \includegraphics[width=\linewidth]{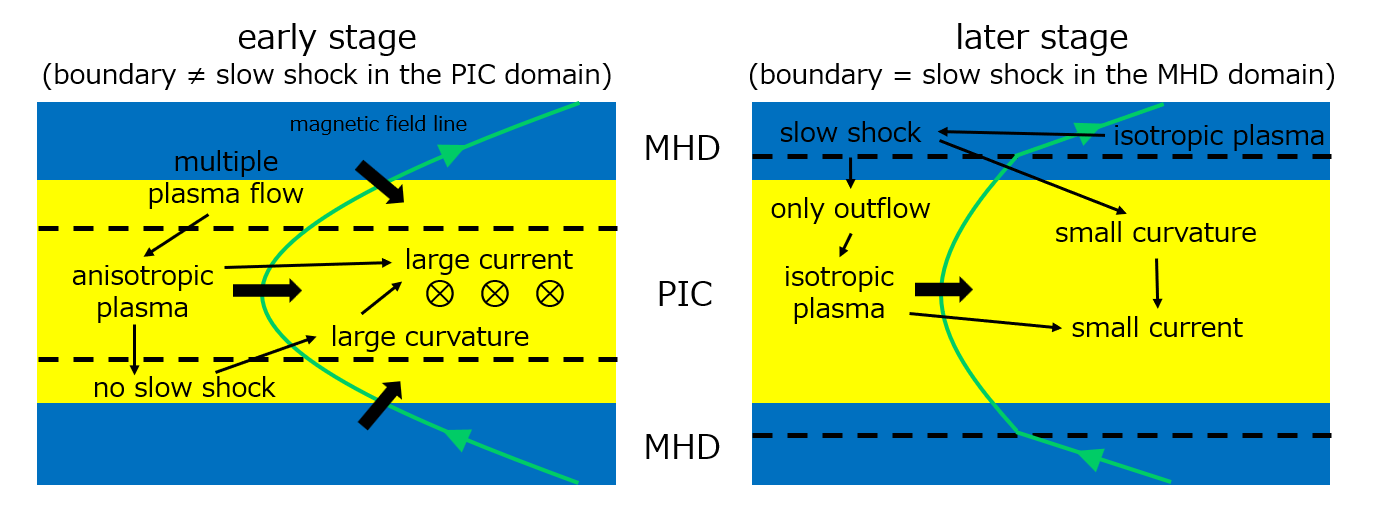}
    \caption{Schematic diagram of multiscale interaction in reconnection exhausts between MHD and PIC domain.}
    \label{chap4:conclusion}
\end{figure*}

In this paper, we performed multi-hierarchy simulation of two-dimensional Riemann problem which models the reconnection outflow region with varying PIC domain sizes to examine the characteristics of the boundary between the inflow and outflow regions. The main findings of this paper are summarized as follows: 
\begin{enumerate}
    \item Although shock is not formed when the boundary between the inflow and outflow regions remains inside the PIC domain, slow shock is generated after the boundary enters the MHD domain. The slow shock is formed and it is nearly the switch-off slow shock ($M_{\rm n1}^2 = 0.7 - 0.8$). 
    \item From RH relations for anisotropic plasmas, slow shocks are rarely formed at the boundary because $\epsilon$ at the downstream is small, making the conditions required to satisfy the RH solution quite stringent. It indicates the boundary between the inflow and outflow regions is not a shock, but a pulse wave. Similar results have been reported in \cite{liu2011pic, higashimori2012}.
    \item Once the slow shock is formed in the MHD domain, $v_y$ decreases so that temperature anisotropy is relaxed by eliminating multiple ion orbits. Around the time when the temperature anisotropy relaxes, the elongated current sheet begins to disappear.  
    \item The slow and intermediate mode waves in anisotropic plasmas with CGL closure have similar group velocity when $\epsilon$ is small. It indicates that the pulse wave behaves a slow and intermediate mode compound wave. Similar results have been reported in \cite{hau1993, liu2011theory}.
    \item When the pulse wave enters the MHD domain, it converts into slow, intermediate, and fast mode wave. Only slow mode wave has large amplitude so that it can steepen to evolve into a slow shock. 
\end{enumerate}
These results suggest that if the mean free path ($\sim$ PIC domain size) is smaller than the entire systems and MHD approximation is valid at a certain scale, nearly switch-off slow shock can be formed in the MHD domain. Moreover, slow shock formation affects the kinetic scale plasma property such as temperature anisotropy. \revise{Our findings suggest that Petschek reconnection in collisionless-collisional systems has multiscale interactions.}{Our findings suggest that Petschek-like reconnection can occur in collisionless–collisional systems, in which temperature anisotropy is relaxed far from the reconnection region by Coulomb collision and MHD approximation is valid, thereby enabling the formation of slow shocks.}

Figure~\ref{chap4:conclusion} presents a schematic summary of the temporal evolution of reconnection exhausts in collisionless--collisional coupled systems. In the early stage, when the boundaries between the inflow and outflow regions remain within the PIC domain, multiple ion orbits generate significant temperature anisotropy \citep{hoshino1998, gosling2005}. This anisotropy inhibits the formation of slow shocks \citep{liu2011pic, liu2011theory, higashimori2012}. In the absence of slow shocks, strong magnetic field curvature develops near the center of the current sheet, resulting in an elongated current layer. The same behavior can also be interpreted as arising from diamagnetic currents associated with the pressure anisotropy \citep{le2016}. In the later stage, when the boundaries move into the MHD domain, slow shocks can form because the MHD assumes isotropic plasma. The slow shocks reduce the inflow velocity and leave only the $v_x$ component in the outflow region, leading to isotropic plasma. The modified magnetic field structure reduces the central field curvature and suppresses the formation of an elongated current sheet. In this regime, the nearly isotropic pressure does not support significant diamagnetic currents. 

\revise{}{In our interpretation, the PIC domain size corresponds to the spatial scale of isotropization due to Coulomb collisions or kinetic effects. In regions located farther from the reconnection region, it is reasonable to assume that the MHD approximation holds due to isotropization processes. The multi-hierarchy model mimics the transition between these regimes without explicitly treating the isotropization processes in the interface domain. In this sense, the model assumes sufficient collisionality in the interface region to couple the MHD and PIC descriptions, and strict consistency between MHD and PIC models is not required. The spatial and temporal scales of isotropization in realistic plasmas remain poorly understood: the spatial scale of solar flares is on the order of $\SI{e7}{\meter}$, the mean free path of Coulomb collisions is $\SI{e5}{\meter}$, and the kinetic scales such as the ion inertial length are as small as $\SI{1}{\meter}$. Resolving this wide range of scales within a collisional PIC model is therefore computationally prohibitively expensive with current resources. This work serves as a proof-of-concept study demonstrating that SSS can be realized when the collisional region is properly taken into account. Verifying the results of this study using a collisional PIC model to assess the impact of artificial isotropization in the interface domain between the MHD and PIC models is important and this will be investigated in future work.}

We discuss why the value of $M_{\rm n1}^2$ is under unity for multi-hierarchy simulations. Figure \ref{chap3:fig:xt_diagram_mode_detection} shows that fast mode wave is generated when the boundary enters the MHD domain. This wave is also seen in Figure \ref{chap3:fig:slice_diagram_magnetic_energy} around $y \sim 200\lambda_i, 300\lambda_i$ at $\Omega_{\rm ci} t = 100.0$. The pulse of fast mode wave steepens and can form fast shock forward and rarefaction wave backward. This rarefaction wave may change the upstream condition of slow shock so that $M_{\rm n1}$ is different from the ideal MHD simulation result.

In the case of the Earth's magnetosphere, the mean free path is sufficiently large compared with the system size that the entire system can be treated as collisionless. The upper and middle panels of Figure \ref{chap3:fig:epsilon_RH_Mn} suggest that the slow shock is hardly formed in the PIC domain, which indicates that Petschek reconnection is less likely to occur in the antiparallel collisionless reconnection. It is suggested that the slow shocks found in the observation of Earth's magnetosphere \citep{saito1995, eriksson2004, walia2024} comes from different physics. Clarifying the reason of slow shock formation in collisionless systems is one of our future works.

On the other hand, for solar flares, the mean free path is $\mathcal{O}(10^5) \lambda_{\rm i}$, whereas the scale of a solar flare is of order $\mathcal{O}(10^{7-8}) \lambda_{\rm i}$. Therefore, solar flares are phenomena in which collisionless and collisional scales interact. Our simulation results show that if the MHD domain is about $10$ times larger than the PIC domain, the slow shock of $M_{\rm n1} \gtrsim \sqrt{0.7} \sim 0.84$ is formed (e.g., for $N_{y, \rm PIC} = 800$ result, PIC domain size is $20 \lambda_{\rm i}$ and the slow shock of $M_{\rm n1}^2 \sim 0.7$ is formed inside $y \sim [200\lambda_{\rm i}, 300\lambda_{\rm i}]$). This indicates that Petschek-like reconnection can be realized in sufficiently large systems compared with the collisionless scale, such as solar flares.

Although the present study is based on a two-dimensional Riemann problem which is thought to describe the situation after the onset of reconnection, rather than a fully self-consistent reconnection simulation, it captures the essential physics of reconnection exhausts and their interaction between MHD and kinetic effects. Performing multi-hierarchy simulations with large simulation box to determine correct reconnection models for solar flares is our future challenge.

\begin{acknowledgments}
Numerical computations were carried out on GPU cluster at the Center for Computational Astrophysics, National Astronomical Observatory of Japan. 
This work was supported by JSPS KAKENHI Grant Numbers JP24K00688, JP25K00976, JP25K01052, and by the grant of Joint Research by the National Institutes of Natural Sciences (NINS) (NINS program No OML032402). 
This work was carried out by the joint research program of the Institute for Space-Earth Environmental Research (ISEE), Nagoya University.
\end{acknowledgments}

\begin{contribution}

Keita Akutagawa: Conceptualization; Data curation; Formal analysis; Investigation; Methodology; Software; Validation; Visualization; Writing – original draft. 
Shinsuke Imada: Conceptualization; Funding acquisition; Supervision; Validation; Writing – review \& editing. 
Munehito Shoda: Supervision; Validation; Writing – review \& editing.


\end{contribution}

%

\software{KAMMUY \citep{akutagawa2025kammuy}}
\software{Thrust \citep{thrust}}
\software{AmgX \citep{amgX}} 
\software{NumPy \citep{numpy}}
\software{Scipy \citep{sciPy}}
\software{Matplotlib \citep{matplotlib}}


\appendix


\section{Asymmetry of the $y - t$ diagram}\label{appendix:asymmetry_of_the_y-t_diagram}

\begin{figure}
    \includegraphics[width=\linewidth]{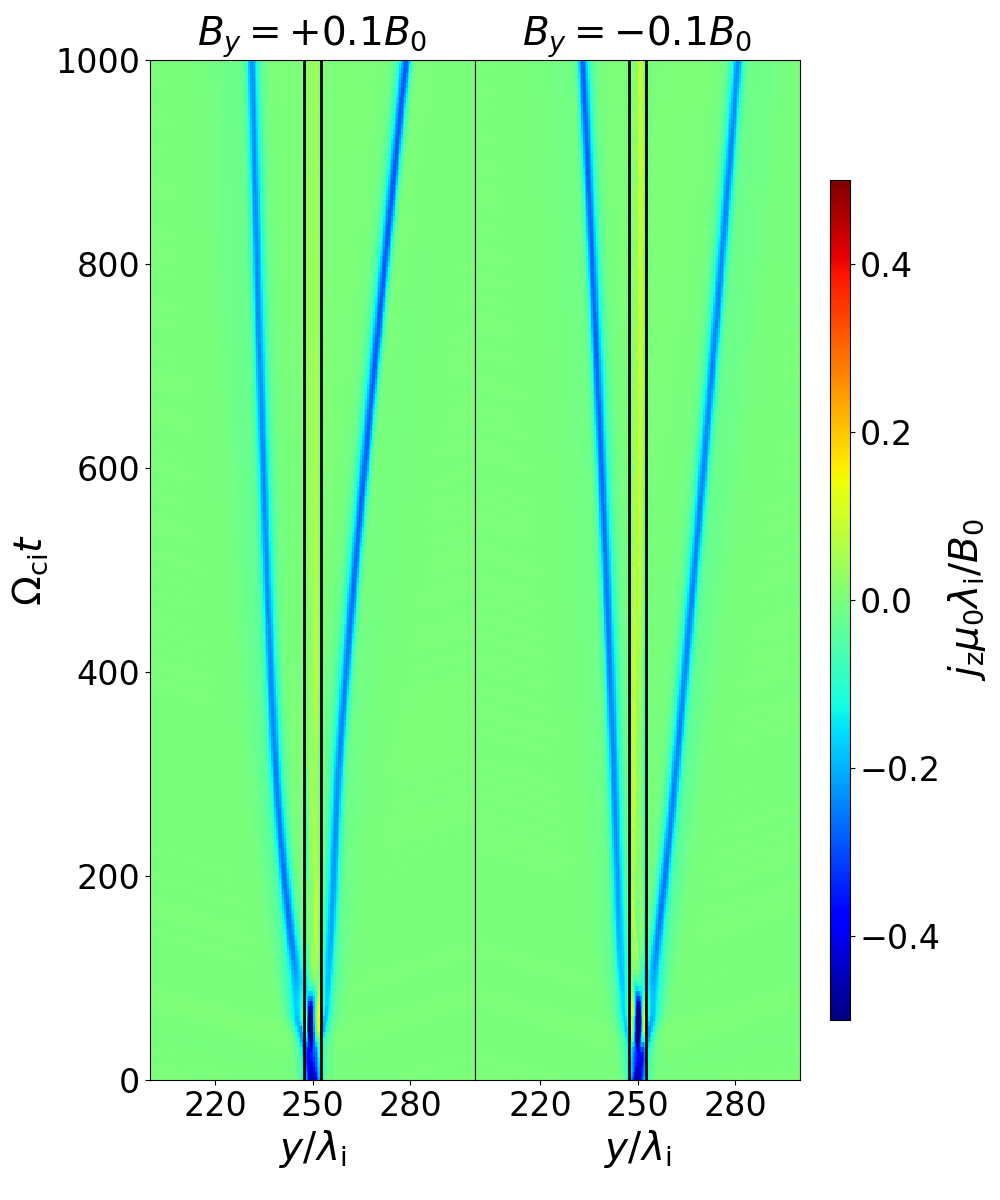}
    \caption{$y - t$ diagrams of $j_{z}$ averaged in the x direction. The panels show the results for $N_{y, \rm PIC} = 100$. $j_{z}$ is normalized by $B_{\rm 0} / \mu_{\rm 0} \lambda_{\rm i}$. The horizontal and vertical axes are normalized by $\lambda_{\rm i}$ and $\Omega_{\rm ci}^{-1}$ respectively.}
    \label{appendix:fig:xt_diagram_jz_minus_by}
\end{figure}

In this appendix, we focus on the asymmetry of the $y - t$ diagram of $j_z$ and $\nabla \cdot \bm{v}$ seen in Figure \ref{chap3:fig:xt_diagram_jz_and_divV}. The asymmetric structure in the initial condition is due to finite $B_y$ component. We performed $B_y = -0.1 B_{\rm 0}$ simulation with $N_{y, \rm PIC} = 100$. Figure \ref{appendix:fig:xt_diagram_jz_minus_by} shows the $y - t$ diagram of $j_z$. At the initial stage ($\Omega_{\rm ci} t \lesssim 400$), the asymmetric structure of $j_z$ is symmetric between $B_y = +0.1 B_{\rm 0}$ and $B_y = -0.1 B_{\rm 0}$. However, at the later stage, the asymmetric structures become similar in both results. It indicates that the asymmetric $j_z$ structure at the initial stage comes from the sign of $B_y$, while that at the later stage the similar structure comes from the different reasons. At the later stage, it is thought that the kinetic effects does not affect the property of slow shocks because they are far from the PIC domain. 


\section{Mach number relation from Rankine--Hugoniot relation of anisotropic plasmas}\label{appendix:anisotropic_plasmas_RH}

In this paper, anisotropic plasma model used in some previous studies \citep{hudson1970, karimabadi1995, higashimori2012} is used to analyze the property of the boundary between the inflow and outflow region. The Rankine--Hugoniot relation from this model is as follows:
\begin{align}
    \left[ \rho v_{\rm n} \right]_2^1 &= 0, 
    \label{appendix:eq:density_jump} \\
    \left[ \rho v_{\rm n^2} + p + \frac{1}{3} \left( \epsilon + \frac{1}{2} \right) \frac{B^2}{\mu_0} - \epsilon \frac{B_{\rm n}^2}{\mu_0} \right]_2^1 &= 0, 
    \label{appendix:eq:normal_moment_jump} \\ 
    \left[ \rho v_{\rm n} v_{\rm t} - \epsilon \frac{B_{\rm n} B_{\rm t}}{\mu_0} \right]_2^1 &= 0, 
    \label{appendix:eq:tangential_moment_jump} \\
    \left[ \rho v_{\rm n} \left( \frac{1}{2} v^2 + \frac{\gamma}{\gamma - 1} \frac{p}{\rho} \right) + \frac{\epsilon + 2}{3} v_{\rm n} \frac{B^2}{\mu_0} \right. \notag \\ 
    \left. - \epsilon v_{\rm n} \frac{B_{\rm n}^2}{\mu_0} - \epsilon v_t \frac{B_{\rm n} B_{\rm t}}{\mu_0} \right]_2^1 &= 0, 
    \label{appendix:eq:energy_jump} \\
    \left[ v_{\rm n} B_{\rm t} - v_{\rm t} B_{\rm n} \right]_2^1 &= 0, 
    \label{appendix:eq:faraday_jump} \\
    \left[ B_{\rm n} \right]_2^1 &= 0,
    \label{appendix:eq:normal_b_jump}
\end{align}
where $[Q]_2^1$ denotes $Q_2 - Q_1$. The subscripts $1$ and $2$ denote the upstream and downstream respectively and the subscripts $\rm t$ and $\rm n$ denote the tangential and normal component to the discontinuity. The pressure $p$ is determined as $p = (p_\parallel + 2 p_\perp) / 3$.

From the definitions of $M_{\rm n} := v_{\rm n} / \sqrt{\epsilon} V_{\rm An}$ where $V_{\rm An} := B_{\rm n} / \sqrt{\mu_0 \rho}$, we get 
\begin{align}
    \rho v_{\rm n}^2 = \frac{\epsilon B_{\rm n}^2 M_{\rm n}^2}{\mu_0}, 
    \label{appendix:eq:rho_vn^2}
\end{align}
and using Equation~\eqref{appendix:eq:density_jump} and ~\eqref{appendix:eq:normal_b_jump}, 
\begin{align}
    \frac{v_{\rm n2}}{v_{\rm n1}} = \frac{\epsilon_2 M_{\rm n2}^2}{\epsilon_1 M_{\rm n1}^2}. 
    \label{appendix:eq:vn2_vn1}
\end{align}

From Equation~\eqref{appendix:eq:tangential_moment_jump}, ~\eqref{appendix:eq:normal_b_jump} and ~\eqref{appendix:eq:rho_vn^2}, we get 
\begin{align}
    \tan \theta_2 = \frac{\epsilon_1 (M_{\rm n1}^2 - 1)}{\epsilon_2 (M_{\rm n2}^2 - 1)}
    \label{appendix:eq:tan_theta2}
\end{align}

Moving into the deHoffman-Teller frame, the velocity is parallel to the magnetic field so that 
\begin{align}
    v_{\rm t} = v_{\rm n} \tan \theta, B_{\rm t} = B_{\rm n} \tan \theta. 
    \label{appendix:eq:vt_and_bt}
\end{align}

From Equation~\eqref{appendix:eq:normal_moment_jump}, ~\eqref{appendix:eq:rho_vn^2} and ~\eqref{appendix:eq:vt_and_bt}, we get 
\begin{align}
    \tilde{\beta}_{\rm 2n} &= \tilde{\beta}_{\rm 1n} + \epsilon_1 M_{\rm n1}^2 + \frac{1}{3 \cos^2 \theta_1} \left( \epsilon_1 + \frac{1}{2} \right) - \epsilon_1 \notag \\
    &-  \epsilon_2 M_{\rm n2}^2 - \frac{1}{3 \cos^2 \theta_2} \left( \epsilon_2 + \frac{1}{2} \right) + \epsilon_2, 
    \label{appendix:eq:beta}
\end{align}
where $\tilde{\beta}_{\rm n} := \mu_0 p / B_{\rm n}^2$. 

From Equation~\eqref{appendix:eq:energy_jump}, ~\eqref{appendix:eq:rho_vn^2}, ~\eqref{appendix:eq:vn2_vn1}, ~\eqref{appendix:eq:tan_theta2}, ~\eqref{appendix:eq:vt_and_bt}, and ~\eqref{appendix:eq:beta}, we get quadratic equation $A (M_{n1}^2)^2 + B M_{n1}^2 + C = 0$ and coefficients are as follows: 
\begin{align*} 
    A &= \frac{\epsilon_1^2}{2 \cos^2 \theta_1} - \frac{\epsilon_2^2 M_{\rm n2}^4 \alpha}{2} - \frac{2}{3} (1 - \epsilon_2) \epsilon_2 M_{\rm n2}^2 \alpha \notag \\ 
    &+ \frac{\gamma}{3(\gamma - 1)} \epsilon_2 M_{n2}^2 \left( \epsilon_2 + \frac{1}{2} \right) \alpha, \\
    B &= \frac{\gamma}{\gamma - 1} \tilde{\beta}_{\rm n1} \epsilon_1 + \frac{2 (1 - \epsilon_1)}{3 \cos^2 \theta_1} \epsilon_1 + \epsilon_2^2 M_{\rm n2}^4 \alpha \notag \\ 
    &- \frac{2 \gamma}{3(\gamma - 1)} \epsilon_2 M_{\rm n2}^2 \left( \epsilon_2 + \frac{1}{2} \right) \alpha \notag \\ 
    &- \frac{\gamma}{\gamma - 1} \epsilon_2 M_{\rm n2}^2 \epsilon_1 + \frac{4 (1 - \epsilon_2)}{3} \epsilon_2 M_{\rm n2}^2 \alpha, \\ 
    C &= -\frac{\epsilon_2^2 M_{\rm n2}^4}{2} (1 + \alpha) - \frac{\gamma}{\gamma - 1} \epsilon_2 M_{\rm n2}^2 \Lambda \notag \\
    &- \frac{2 (1 - \epsilon_2)}{3} \epsilon_2 M_{\rm n2}^2 (1 + \alpha), \\
    \label{appendix:eq:Mn1_eq}
\end{align*}
where 
\begin{align*}
    \Lambda &= \tilde{\beta}_{\rm n1} + \frac{1}{3 \cos^2 \theta_1} \left( \epsilon_1 + \frac{1}{2} \right) - \epsilon_1  \notag \\
    &- \epsilon_2 M_{\rm n2}^2 - \frac{1}{3} \left( \epsilon_2 + \frac{1}{2} \right) (1 + \alpha) + \epsilon_2, \\
    \alpha &= \epsilon_1^2 \tan^2 \theta_1 / \epsilon_2^2 (M_{\rm n2}^2 - 1)^2.
\end{align*}


\bibliography{bib}
\bibliographystyle{aasjournalv7}



\end{document}